\documentclass[reprint, amsmath, amssymb, prb, aps, superscriptaddress]{revtex4-1}

\usepackage{graphicx}
\usepackage{bm}
\usepackage{subfig}
\usepackage{caption}
\usepackage{amsmath}
\usepackage{color}
\usepackage{appendix}
\usepackage[normalem]{ulem}

\begin{document}

\title{Determination of Fermi surface by charge density correlations}

\author{Zhipeng Sun}%
\email{zpsun@csrc.ac.cn}
\affiliation{Beijing Computational Science Research Center, Beijing 100193, China}%

\date{\today}

\begin{abstract}
    The Fermi surface topology in the two-dimensional Hubbard model is particularly relevant for the high-temperature superconductors, whereas its theoretical research encounters with the difficulty of the analytical continuation problem. To this end, we proposed the concept of the momentum-dependent compressibility, defined as the variation of the momentum distribution function with respect to the chemical potential. The surface determined by the maximum of the momentum-dependent compressibility is nearly identical to the Fermi surface in the weakly and intermediate coupling regions according to our numerical results. In the correlated region, this surface also exhibits pocket and arc features, just like the Fermi surface in high-temperature superconductors. Therefore, for theoretical studies, this surface can be used as an alternative to determine the underlying Fermi surface. Considering that the momentum-dependent compressibility is closely related to the charge density correlations, our work also shows a connection between the Fermi surface topology and the charge density fluctuations.

\end{abstract}

\maketitle

\textit{Introduction}. --- Over the past century, the notion of the Fermi surface (FS) has played an central role in our understanding of the physical properties of solid-state materials. For conventional metals that can be described by Landau's Fermi liquid theory, the FS can be obtained from the local density approximation calculations, and the results are in good agreement with the experiments. However, in the high-temperature superconductors (HTSC), the FS is not well defined due to the failure of the quasiparticle picture \cite{norman1998}. Instead, the Luttinger surface \cite{dzyaloshinskii2003}, defined through Luttinger's theorem, is taken as the underlying FS \cite{gros2006}, also commonly referred to as the FS. The determination of the FS in experiments is based on the data measured by the angle-resolved photoemission spectroscopy (ARPES) \cite{PhysRevB.63.224516, damascelli2003, sobota2021}, and the FS of HTSC exhibits the pocket and arc features \cite{PhysRevLett.95.077001, tanaka2006}. These features are naturally linked to the strong correlated effects such as the pseudogap and the superconductivity \cite{PhysRevLett.114.147001, doiron2017, vishik2018, mitsen2019}, but their relationship remains a mystery.

In theoretical study, the FS in the correlated electronic models is usually determined by the zero-frequency spectral intensity \cite{PhysRevB.66.075102, PhysRevX.8.021048}, which is obtained through the analytical continuation procedure \cite{PhysRevB.81.155107, PhysRevLett.126.056402} based on the data of the Matsubara Green's function. However, the analytical continuation is an ill-conditioned problem, which is usually very sensitive to the precision of the data, and thus the results are not convincing enough. Therefore, a method to determine the underlying FS without the analytical continuation is highly desirable for theorectical researchers.

In this paper, we proposed the notion of the momentum-dependent compressibility, defined as the variation of the momentum distribution function with respect to the chemical potential. Our theoretical analysis in the non-interacting case shows that the maximum of the momentum-dependent compressibility constitutes a surface that is consistent with the FS determined by the spectral intensity. We also carried out numerical calculations in the two-dimensional Hubbard model to numerically illustrate the similarity of this surface to the FS in the weak and  intermediate coupling cases. The surface also exhibits the pocket and arc features in the strongly correlated region, just like the FS in HTSC.

\textit{Notion of momentum-dependent compressibility}. --- Recall the ideal electronic gas system at finite temperature with dispersion $\varepsilon_{\bm{k}}$. The momentum distribution function is $n_{\bm{k}} = \frac{2}{\mathrm{e}^{\beta \left(\varepsilon_{\bm{k}}-\mu\right)} + 1}$ with $\beta$  the inverse temperature and $\mu$ the chemical potential. The FS can then be determined by the equation $n_{\bm{k}}=1$. Note that the variation of the momentum distribution function with respect to the chemical potential is equals to
\begin{equation}
    \frac{dn_{\bm{k}}}{d\mu} = - \frac{1}{2}\beta n_{\bm{k}}^{2} + \beta n_{\bm{k}}. \label{eq:dnk-gas}
\end{equation}
Obviously, the quantity $dn_{\bm{k}}/d\mu$ reaches its maximum at where $n_{\bm{k}}=1$. That is, in the noninteracting cases, the surface determined by $dn_{\bm{k}}/d\mu$ is identical to the FS. 

The quantity $dn_{\bm{k}}/d\mu$ has its own physical meanings. Firstly, it described the change rate of the momentum distribution function $n_{\bm{k}}$, and the Fermi momentum is just where the occupance varies the fastest. Secondly, it is a component of the charge compressibility $\chi^{\text{c}}\equiv dn/d\mu$ with $\mu$ the average charge density, therefore, we call it as the momentum-dependent compressibility. Thirdly, $d_{\bm{k}}/d\mu$ relates to the charge density correlations:
\begin{equation}
    \begin{aligned}
        \frac{dn_{\bm{k}}}{d\mu} &= \sum_{\bm{k}^{\prime}} \int_{0}^{\beta} d\tau^{\prime} \ 
        \left\langle n_{\bm{k}}\left(\tau=0\right)  n_{\bm{k}^{\prime}}\left(\tau^{\prime}\right) \right\rangle \\ &\quad
        - \left\langle n_{\bm{k}}\left(\tau=0\right) \right\rangle 
        \sum_{\bm{k}^{\prime}}\int_{0}^{\beta} d\tau^{\prime} \ 
        \left\langle n_{\bm{k}^{\prime}}\left(\tau^{\prime}\right) \right\rangle. 
    \end{aligned}
\end{equation}
It stands for the correlation between the ${\bm{k}}$-specified charge density and the total charge density. 

\textit{Model and method}. --- We study the surface determined by $dn_{\bm{k}}/d\mu$ in the two-dimensional square Hubbard model using the $HGW$ method (see Ref. \onlinecite{PhysRevB.104.125137} or the Supplemental Material \footnote{\label{sm}See Supplemental Materials for details of the $HGW$ formalism and more supporting numerical results.}). The Hamiltonian reads
\begin{equation}
    \mathcal{H} = -\sum_{ij\sigma} t_{ij} \hat{c}_{i\sigma}^{\dagger} \hat{c}_{j\sigma} + \frac{U}{2} \sum_{i\sigma} \hat{n}_{i\sigma} \hat{n}_{i\bar{\sigma}} - \mu \sum_{i\sigma} \hat{n}_{i\sigma}.
\end{equation}
Here $\hat{c}_{i\sigma}^{\dagger}$ ($\hat{c}_{i\sigma}$) is the creation (annihilation) operator for the electron with spin $\sigma$ at lattice site $i$, and $\hat{n}_{i\sigma}\equiv \hat{c}_{i\sigma}^{\dagger} \hat{c}_{i\sigma}$ is the density operator. $t_{ij}$ is hopping strength for the electron from site $i$ to site $j$, $U$ is the on-site repulsive interaction, and $\mu$ is the chemical potential. Here we only consider the nearest-neighbor hopping strength $t$ and the next nearest-neighbor hopping strength $t^{\prime}$, and set $t=1$ as the unit of energy. 

The $HGW$ method is based on the truncation of the correlation functions in the equation of motions, and the resulting equations are referred as the $HGW$ equations. By solving the $HGW$ equations, one obtains the Matsubara Green's function, and then through the Nevanlinna analytical continuation \cite{PhysRevLett.126.056402}, one obtains the spectral function and thus the FS. 

Restricted by the fluctuation-dissipation theorem, the two-body correlations are calculated through the functional derivative scheme within the $HGW$ equations. As for the momentum-dependent compressibility $dn_{\bm{k}}/d\mu$, one can also, in principle, calculate it from independent calculations of the momentum distribution function $n_{\bm{k}}$ for different values of the chemical potential. It is worth to note that these two calculations are consistent as shown in Ref. \onlinecite{PhysRevB.104.125137} due to the respecting of the fluctuation-dissipation theorem.

\begin{figure}[!h]
    \centering
    \includegraphics[width=0.95\linewidth]{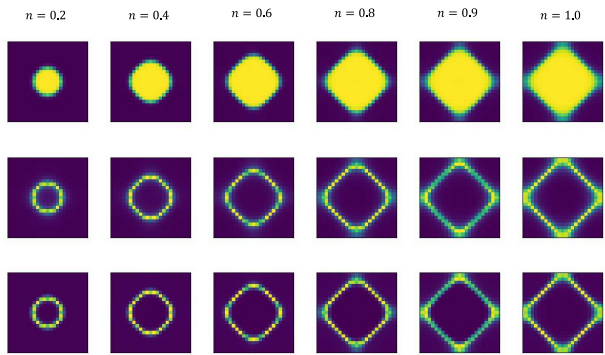}
    \captionsetup{font={small},justification=raggedright}
    \caption{The plot of momentum distribution function $n_{\bm{k}}$ (top panel), zero-frequency spectral intensity $A\left(\bm{k}, \omega=0\right)$ (medium panel) and momentum-dependent compressibility $dn_{\bm{k}}/d\mu$ (bottom panel) at weak coupling $U=2.0$. From left to right, the fillings are $0.2$, $0.4$, $0.6$, $0.8$, $0.9$, $1.0$. \label{fig:u2}}
\end{figure}

\begin{figure}[!h]
    \centering
    \includegraphics[width=0.95\linewidth]{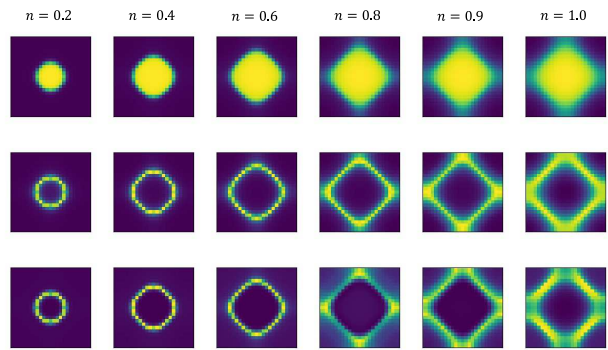}
    \captionsetup{font={small},justification=raggedright}
    \caption{The plot of momentum distribution function $n_{\bm{k}}$ (top panel), zero-frequency spectral intensity $A\left(\bm{k}, \omega=0\right)$ (medium panel) and momentum-dependent compressibility $dn_{\bm{k}}/d\mu$ (bottom panel) at intermediate coupling $U=4.0$. From left to right, the fillings are $0.2$, $0.4$, $0.6$, $0.8$, $0.9$, $1.0$. The temperature is $1/5.0$ for $n=1.0$ and $1/8.0$ for other fillings. \label{fig:u4}}
\end{figure}

\textit{Numerical results.}. --- Consider the weak coupling $U=2.0$ and intermediate coupling $U=4.0$ cases with $t^{\prime}=0$ at the temperature $T=1/8.0$, for different charge fillings on a $24\times 24$ lattice, which can be compared with the results obtained by quantum Monte Carlo simulations in Ref. \onlinecite{PhysRevB.80.075116}. The plot of the momentum distribution function $n_{\bm{k}}$, the zero-frequency spectral intensity $A\left(\bm{k}, \omega\right)$ and the momentum-dependent compressibility $dn_{\bm{k}}/d\mu$ is shown in Fig. \ref{fig:u2} for $U=2.0$ and Fig. \ref{fig:u4} for $U=4.0$. The results show the two surfaces determined by $A\left(\bm{k}, \omega\right)$ and $dn_{\bm{k}}/d\mu$ are almost identical. Noting that the dinotomy of the FS near the antinode $\left(\pi, 0\right)$ indicates the precusor of the pseudogap, the surface determined by $dn_{\bm{k}}/d\mu$ also displays a similar dinotomy at and near half-filling.

\begin{figure}[!h]
    \centering
    \includegraphics[width=0.9\linewidth]{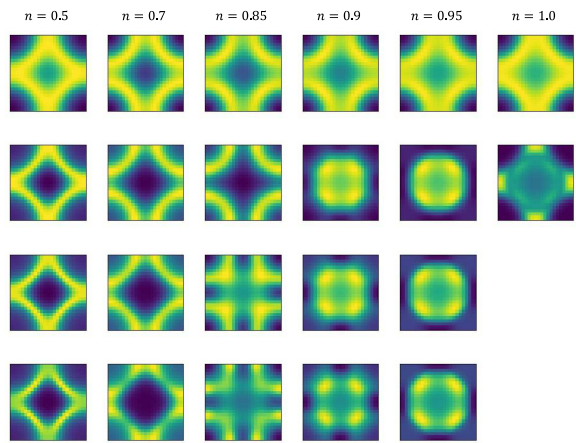}
    \captionsetup{font={small},justification=raggedright}
    \caption{The plot of $dn_{\bm{k}}/d\mu$ for $U=6.0$ and $t=-0.25$ for $24\times24$ lattice. From left to right, the charge fillings are $0.5$, $0.7$, $0.85$, $0.9$, $0.95$, $1.0$. From top to bottom, the inverse temperatures are $1.0$, $3.0$, $5.0$ and $8.0$. \label{fig:u6}}
\end{figure}

Then we choose $U=6.0$ and $t^{\prime}=-0.25$ representing for the parameter for the correlated materials, used in Ref. \onlinecite{huang2019}. The plot of $dn_{\bm{k}}/d\mu$ for different fillings and different temperatures is shown in Fig. \ref{fig:u6}. The results show that the surface exhibits the large pocket (in the overdoped region $n\le 0.85$) and small arc (in the underdoped region $n=0.9,\ 0.95$) features, just like the FS does in the HTSC \cite{PhysRevB.74.224510, drozdov2018}. Remarkably, the surface at $n=1.0,\ 1/T=3.0$ where the system is nearly in the Mott phase qualitatively matches with the remnant Fermi surface of the Mott insulator $\text{Ca}_{2}\text{Cu}\text{O}_{2}\text{Cl}_{2}$ \cite{ronning1998}. 

\begin{figure}[!h]
    \centering
    \includegraphics[width=0.9\linewidth]{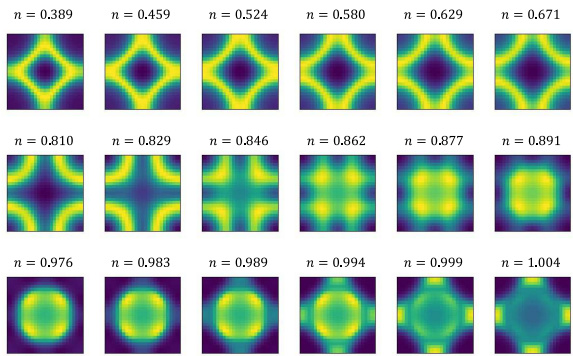}
    \captionsetup{font={small},justification=raggedright}
    \caption{The change process of the surface topology slowly with the filling number for fixed temperature $1/T=3.0$. The top panel shows the change process of the surface from closed to open, the middle panel shows the process that the surface changes form pocket to arc, and the bottom panel shows the process of surface changing from arc to Mott feature. \label{fig:st}}
\end{figure}

We further observed that the surface topology changes slowly with the filling number, with the fixed temperature $1/T=3.0$. As shown in the Fig. \ref{fig:st}, the top panel shows the change process of the surface from closed to open (hole pocket), namely the so-called Lifshitz transition. At very small filling, for example $n=0.389$, the suface is in a closed form. Near the quarter filling, $n=0.524$, the surface meets the antinode $\left(\pi,0\right)$. As the filling continues to increase, the surface changes to an open form. The middle panel shows the process that the surface changes form pocket to arc. In the overdoped region, for example $n=0.810$, the surface is still in an open form, which is also referred as the ``hole pocket''. Near the critical doping, $n=0.846$, the value of $dn_{\bm{k}}/d\mu$ starts to increase (relatively to the compressibility $\chi^{\text{c}}$) at near node $\left(\pi/2, \pi/2\right)$ and decrease at near antinode (along $\left(\pi,0\right)$ to $\left(\pi,\pi\right)$). And then in the underdoped region, the surface exhibits an arc feature. The bottom panel shows the process of surface changing from arc to Mott feature where a small pocket forms near the antinode. 

\begin{figure}[!h]
    \centering
    \includegraphics[width=0.9\linewidth]{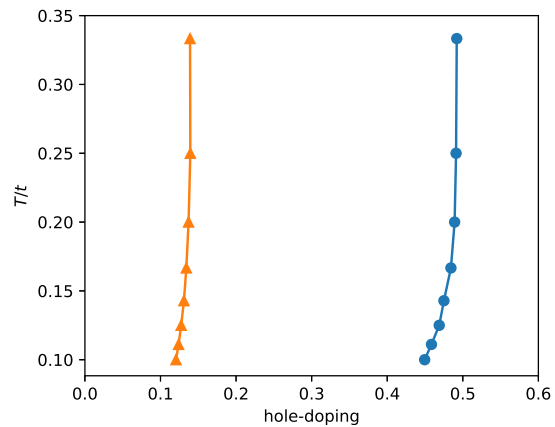}
    \captionsetup{font={small},justification=raggedright}
    \caption{The phase diagram determined by the surface topology for $U=6.0$ and $t=-0.25$ for $24\times24$ lattice, relating to the Fermi liquid (right region), the strange metal state (middle region), and the pseudogap region (left region) in the cuprate superconductors. \label{fig:pd}}
\end{figure}

Fig. \ref{fig:pd}  shows the phase diagram determined by the surface topology. The blue line is the boundary between the closed surface and the open surface, and the orange line distincts the pocket and the arc \footnote{The blue line is where the maximum value of $n_{\bm{k}}$ along $\left(0,0\right)$ to $\left(\pi,0\right)$ equals to the maximum value along $\left(\pi,0\right)$ to $\left(\pi, \pi\right)$. The orange line is where the maximum value along $\left(\pi,0\right)$ to $\left(\pi,\pi\right)$ equals to $0.8$ times the global maximum. }. Compared with the phase diagram of cuprate superconductors\cite{sobota2021}, it is not difficult to speculate that the closed surface corresponds to the Fermi liquid, the open surface stands for the strange metal state, and the arc implies the pseudogap.  

\textit{Summary and discussion}. --- In summary, we proposed the notion of the momentum-dependent compressibility. The surface determined by the maximum of this notion is closely related to the FS. Our numerical results show that, this surface is almost identical to the FS in the weak coupling and intermediate coupling regions. In the correlated region, this surface also exhibits the pocket and arc features, just like the FS does in HTSC. 

It is worth emphasizing that the surface determined by the momentum-dependent compressibility is a well-defined concept in general cases. Therefore, it can be an alternative to determine the underlying FS for theorectical researches. Especially, the calculation for this surface does not require the analytical continuation, which brings great convenience for the implementation. 

Note that the momentum-dependent compressibility is closely related to the charge density correlations. Its behavioral similarity to the zero-frequency spectral function indicates the connection between the charge density fluctuations and the FS topology. Therefore, the understanding of the FS topology promotes the understanding of the charge density fluctuations. 

\bibliography{reference}

\end{document}